# A PMUT Integrated Microfluidic System for Fluid Density Sensing

Kaustav Roy, *Student Member, IEEE*, Kritank Kalyan, Anuj Ashok, Vijayendra Shastri, Antony Jeyaseelan, Avinandan Mandal and Rudra Pratap, *Senior Member, IEEE*

*Abstract*— We demonstrate the design, fabrication and use of a dual electrode PMUT (Piezoelectric Micromachined Ultrasound Transducer) integrated with a microfluidic channel as a fluid density sensor in both static and dynamic density-change conditions. The dual electrode configuration makes the PMUT resonator a self-contained resonant peak-shift sensor and the microfluidic integration makes this system a versatile fluid density sensing platform that can be used with extremely low volumes of fluids in various industrial and healthcare applications. The density measurements carried out here under flowing fluid conditions demonstrate the potential of this system as a real-time fluid density monitoring system. We include results of density measurements in the range of 1020 - 1090 kg/m$^3$ that corresponds to the human blood density variation generally due to the change in its hemoglobin content. The sensitivity of the sensor—26.3 Hz/(kg/m$^3$)—is good enough to reliably detect even 1% change in the hemoglobin content of the human blood. Thus, this system could potentially be used also as a hemoglobin measurement sensor in healthcare applications.

## I. Introduction

FLUIDS have certain physical properties, which, if tracked, can provide information about their quality and condition. Often this information is related to the mechanical properties of the fluid such as density, dynamic viscosity, volume viscosity, bulk modulus, and surface tension. Monitoring these properties is highly desirable in both industrial and healthcare applications. These properties can be monitored in real-time using fluid transducers [1]–[16], which can be fabricated in various sizes with diverse specifications.

We have already worked on realizing such transducers at microscale, and a class of piezoelectric-MEMS based fluid transducers using 2D resonators in the form of piezoelectric micromachined ultrasound transducers (PMUTs) [17]–[20] have been fabricated for density sensing. Such a fluid transducer that used through transmission of ultrasound in an arrangement of PMUT-Fluid-PMUT (PFP) in order to monitor the fluid density in real-time was reported recently [21], [22]. Although there are numerous advantages of this system, it has a serious drawback in terms of its size and complex packaging. Also, it requires a pair of PMUTs, one acting as a transmitter and the other as a receiver in order to determine the density of a fluid. These complications can be avoided by fabricating dual-electrode PMUTs [23], thereby enabling density sensing using a self-sensing mechanism on a single transducer.

We fabricate dual-electrode PMUTs in this work, characterize them, and study their applicability as fluid density sensors. We fabricate a linear array of single-cell PMUTs and integrate them to a PDMS-based microfluidic channel, thereby forming the first successful PMUT-microfluidic-integration (PMI). Such an integrated system works as a self-contained microfluidic density sensing platform capable of real-time fluid density monitoring and is being reported here for the first time as far as we know. We also demonstrate that the dual-electrode single-cell PMUTs are capable of detecting density changes in the range of density variation of blood due to the change in hemoglobin concentration [24], thereby leading to the preliminary development of a blood hemoglobin sensor.

This work involves PMUT design, microfluidic channel design, fabrication of these two components, their characterization, integration of the PMUT with the microfluidic channel, and measurements on fluid samples for density sensing. The rest of the paper contains presentation and discussion in the same order.

## II. Dual Electrode PMUTs As Fluid Density Sensor

### A. Dual Electrode PMUT Topology

The PMUTs used in this work were fabricated at the National nanofabrication Centre, the Centre for Nano Science

This work is submitted for review on 1/03/2021. This work was funded from the grant – STARS/APR2019/NS/653/FS from the Ministry of Human Resource and Development (MHRD)

About authors: Kaustav Roy is with Centre for Nano Science and Engineering, IISc Bangalore, 560012, India (e-mail: kaustav@iisc.ac.in).
Kritank Kalyan is with Centre for Nano Science and Engineering, IISc Bangalore, 560012, India (e-mail: kritankk@iisc.ac.in).
Anuj Ashok is with Centre for Nano Science and Engineering, IISc Bangalore, 560012, India (e-mail: anujashok@iisc.ac.in).

Vijayendra Shastri is with the Centre for Nano Science and Engineering, IISc Bangalore, 560012, India (e-mail: vijayendra@iisc.ac.in).
Antony Jeyaseelan is with the Centre for Nano Science and Engineering, IISc Bangalore, 560012, India (e-mail: antonya@iisc.ac.in).
Avinandan Mandal was with the Centre for Nano Science and Engineering, IISc Bangalore, 560012, India (e-mail: avinandanm@iisc.ac.in).
Rudra Pratap is a joint faculty of Mechanical Engineering Department and Centre for Nano Science and Engineering, IISc Bangalore, 560012, India (email: pratap@iisc.ac.in)

and Engineering, Indian Institute of Science, Bangalore. These devices use lead zirconate titanate (PZT) as the active layer. A PZT thin film of thickness ~ 500 nm was deposited on a platinized silicon-on-insulator (SOI) substrate having a constant device layer thickness of 10 µm. This PZT deposition was done at the National Aerospace Laboratories, Bangalore. The fabricated PMUTs incorporate two active electrodes, namely the driving electrode (D.E.) and the sensing electrode (S.E.), as shown in Fig. 1. The fabricated PMUTs were circular in shape and could be designed to operate at desired frequencies. This is done by either varying the diameter, the thickness of the diaphragm stack, or the interlayer stresses. In this work, the diameter of the PMUTs was varied in order to obtain desired frequencies while keeping all other design variables fixed (thickness of the diaphragm was fixed at ~ 10.5 µm and the interlayer residual stresses at ~ 700 MPa, tensile). There are various topologies and approaches to design and fabricate PMUTs that one can follow [25], and for a desired resonant frequency, there are plenty of geometric design choices.

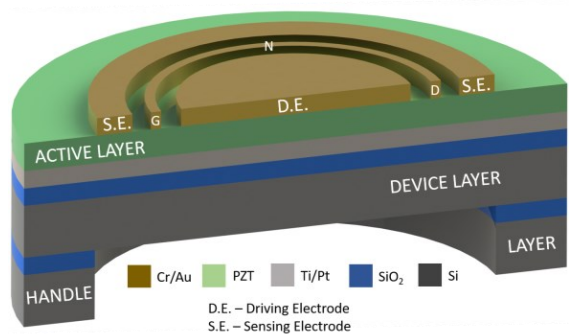

Fig. 1. A 3-D schematic of a single cell dual-electrode PMUT showing the constituent layers

### B. Working Principle

The resonant sensing for fluid density works on a principle called the virtual added mass effect, which refers to the fluid density dependent extra mass that acts on a vibrating object when it vibrates in the fluid of interest. This virtual mass shifts the resonant frequency of the structure in direct relation to the medium density and thus this shift can track even a minute dynamic change in the density of the medium if the change in the resonant frequency can be tracked at a much faster time scale than the change in density. A denser medium causes a decrease in the resonant frequency of the structure, and a rarer medium increases the frequency, (for example, the frequency obtained from a 750 µm PMUT in carbon tetrachloride, having 1496 kg/m$^3$ is ~ 86 kHz, whereas the frequency obtained from ethanol, having density 774 kg/m$^3$ is ~ 65 kHz) thereby making the resonators a preferred candidate for density sensing. At microscales, one of the most suitable candidates for resonant density sensing is a PMUT, which is in essence a 2D microplate resonator. In this work, single-cell PMUTs have been fabricated having two electrodes, one circular central electrode covering 70% of the PMUT area and the other annular electrode placed 20 µm away from the central electrode towards the fixed edge. The central and the annular electrodes are separated by an electrical ground line, GND, to eliminate the effect of cross-coupling (see Fig. 1). The central electrode is used as the driving electrode, and the annular electrode is used as the sensing electrode. To actuate the PMUT, an AC voltage of constant magnitude is applied to the driving electrode while sweeping the signal frequency over the desired range, which, in turn, vibrates the PMUT diaphragm. The peak displacement of the diaphragm is recorded at its resonant frequency, which depends on the medium density. The corresponding strain developed in the diaphragm gives rise to a proportionate voltage output from the sensing electrode through the direct piezoelectric effect. This output voltage is plotted against the swept frequency to generate a frequency response function that contains the resonant frequency corresponding to the density of the medium.

### C. Fabrication of PMUTs

PMUTs used in this work were fabricated using the process flow shown in Fig. 2. The process starts with a platinized

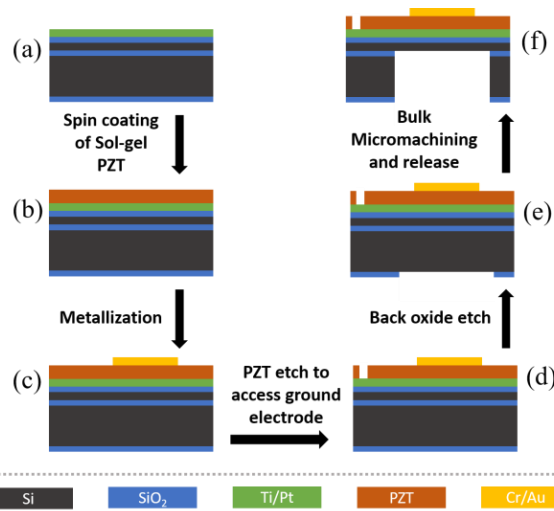

Fig. 2. Process flow describing the steps involved in the fabrication of PMUT (a) The process starts with a platinized SOI wafer (b) PZT sol is spin coated on the wafer (c) The wafer is metallized to create the top electrode contact (d) PZT (d) PZT is etched to access the bottom electrode contact (e) Backside oxide is patterned and etched by RIE (f) The bulk silicon is etched using DRIE

silicon-on insulator wafer having a 10 µm device layer (Fig. 2(a)). Lead zirconate titanate (PZT) is spin-coated repeatedly using the sol-gel technique to achieve the desired thickness (~

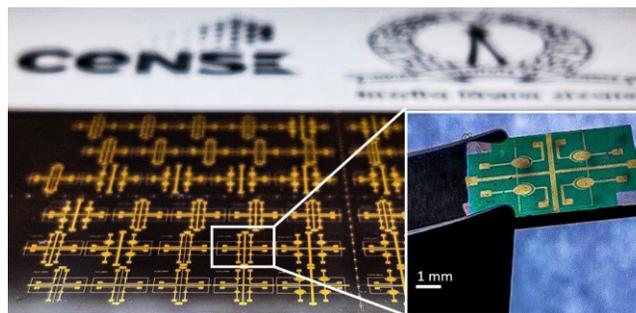

Fig. 3. An array of dual-electrode PMUT devices with a zoomed-in view of a typical dual-electrode PMUT die containing four single cell PMUTs

0.5 µm) and then annealed at 650°C, leading to the formation of a uniform thin film (Fig. 2(b)). The top electrode (Cr/Au – 30/120 nm) is patterned using lithography (Fig. 2(c)) followed by wet etching of the PZT layer (Fig. 2(d)). The stack is then patterned from the backside to etch the backside oxide using reactive ion etching (RIE) as shown in Fig. 2(e). Finally, the handle layer is removed using deep reactive ion etching (DRIE) in order to release the devices, followed by the etching of the buried oxide in order to make the released stack stress-free (Fig. 2(f)). The dual electrode PMUTs fabricated by the above-mentioned process flow is shown in Fig. 3.

### D. Characterization of PMUTs

#### 1) Material Characterization

In order to assess the quality of the PZT thin film, it was characterized using a scanning electron microscope (SEM). The film was found to be free of defects and uniformly spread over the Ti/Pt thin film. The thickness of the PZT film was found to be 466.67 nm (see Fig. 4(a)). To assess the orientation of the deposited PZT film, X-ray diffraction scan was carried out. The θ-2θ plot obtained is shown in Fig. 4(b). The peaks are indexed as per the JCPDS card no: 33-0784 which confirms the tetragonal perovskite structure of the PZT film. The deposited film showed the most prominent peak along (111) orientation. This indicates that the highly oriented bottom Pt (111) electrode acts as a nucleation center for a favored orientation of the PZT thin film. This preferred orientation influences its electrical and mechanical performance.

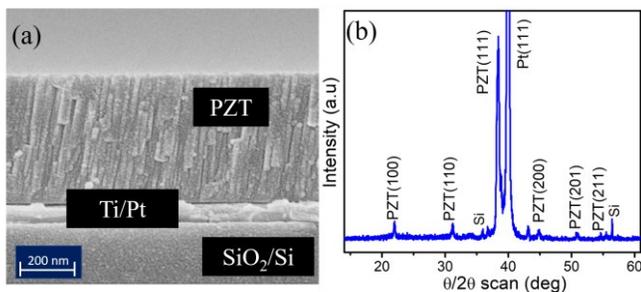

Fig.4. (a) Scanning electron micrograph (SEM) of the cross section of the layered stack in a typical PMUT. (b) θ-2θ scan of the thin film stack used to make the PMUT, showing clearly the (111) oriented peak of PZT

#### 2) Electrical Characterization

To ascertain the quality of the thin-film PZT electrically, polarization was done by sweeping the applied voltage using the Precision Materials Analyzer from Radiant Technologies

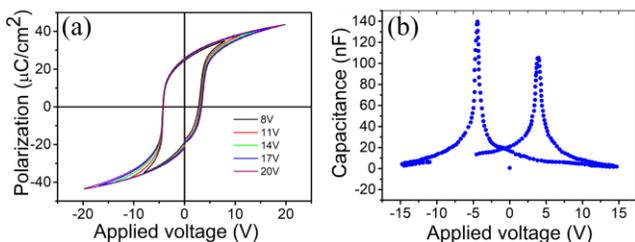

Fig. 5. (a) Polarization vs. applied voltage showing the ferroelectric hysteresis loop. (b) Capacitance vs. applied voltage showing the hysteresis

Inc., and the hysteresis loop so obtained is shown in Fig. 5(a). It shows a typical ferroelectric switching behavior of the thin film PZT with the applied electric field. The obtained remanent polarization ($P_r$), saturated polarization ($P_s$), and the coercive field ($E_c$) from the plot are found to be 26 µC/cm$^2$, 44 µC/cm$^2$, and 3.02 V, respectively. Capacitance is also plotted with respect to the applied voltage, and a hysteresis plot is observed as shown in Fig. 5(b), which suggests the ferroelectric switching phenomenon. The peak capacitances were observed to be 140.4 nF and 105.3 nF, respectively. The asymmetry in the peak capacitances can be due to the lead vacancies in the deposited PZT film as reported in the literature [26]. To understand the AC characteristics of the film, electrical characterization was conducted using 4294A Precision Impedance Analyzer, Agilent Inc., and the results obtained are shown in Fig. 6. The frequency was swept from 230 kHz to 245 kHz. The peak impedance was found to be 101.8 Ω and the peak capacitance was calculated to be 6.7 nF.

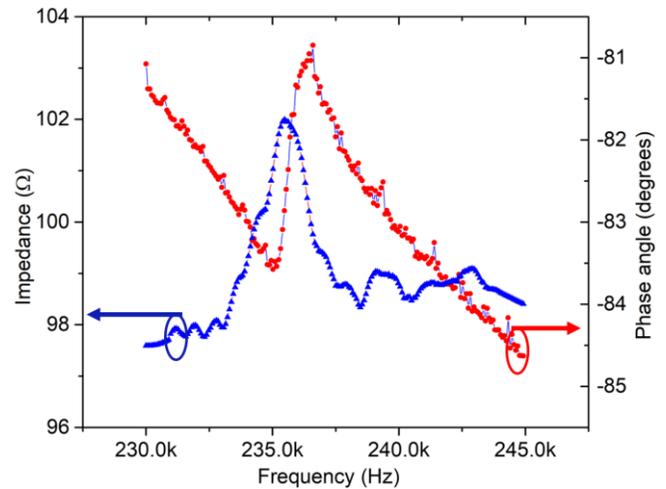

Fig. 6. Impedance and phase angle vs. swept frequency as obtained from the impedance analyzer in air for a typical 750 µm diameter PMUT

#### 3) Electro-Mechanical Characterization

PMUTs of two different sizes were fabricated and characterized for vibration response using a laser Doppler vibrometer (Micro System Analyzer, MSA 500 from Polytec

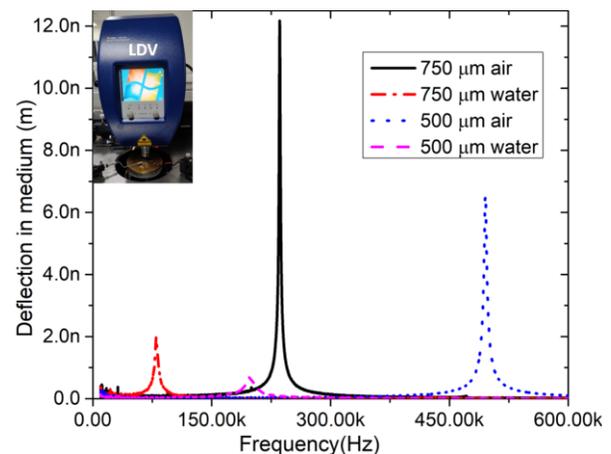

Fig. 7. Frequency response of dual electrode PMUTs in air and water

Inc.) The resonant frequencies of 750 µm and 500 µm diameter PMUTs were found to be 235.1 kHz and 496.8 kHz in air, and 79.2 kHz and 196 kHz in water, respectively, as shown in Fig.

7. In order to find the actual deflection of the PMUTs and their working as an actuator at their resonance, a frequency sweep was carried out over an appropriate frequency band using the peak hold settings and by actuating the driving electrode with 0.5V AC. Similarly, experiments were carried out using a lock-in amplifier (MFLI) from Zurich Instruments to verify the performance of the PMUT as a sensor. The experiments were carried out in air. The response obtained from the LDV and the lock-in amplifier are tabulated in Table I.

TABLE I
ELECTROMECHANICAL RESPONSE OF DUAL ELECTRODE PMUTs

| Diameter (μm) | Resonant Frequency in air (kHz) | Peak deflection measured with LDV. (μm) | Peak voltage sensed using lock-in amplifier (mV) |
|---|---|---|---|
| 750 | 235.1 | 1.5 | 8.6 |
| 500 | 496.8 | 0.85 | 3.3 |

III. PMUT MICROFLUIDIC INTEGRATION (PMI)

In order to fabricate the microfluidic density sensor, it was necessary to create a microfluidic channel and integrate it with PMUTs. This was done by fabricating a single PDMS channel having an inlet and an outlet, aligning, and bonding it to a specially designed PMUT array containing three PMUTs as shown Fig. 8. For convenience, we name this device PMUT-Microfluidic-Integration (PMI). It forms a self-sensing platform to monitor fluid density in microfluidic regimes.

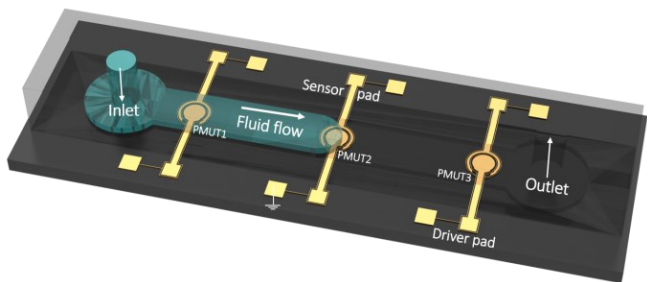

Fig. 8. A 3-D schematic of the PMUT-Microfluidic-Integration (PMI)

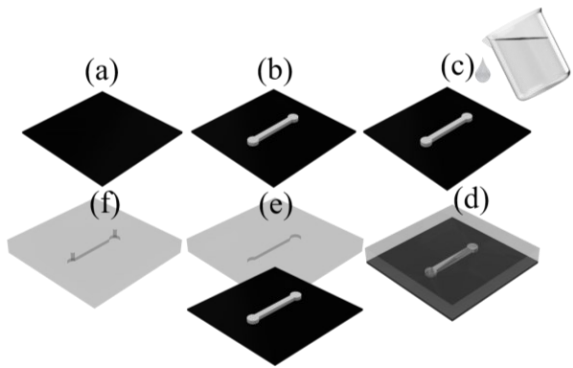

Fig. 9. A 3-D schematic of the fabrication process flow for fabricating PMI (a) Process starts with a silicon wafer (b) Patterning and timed etching of the wafer by DRIE to form the microfluidic mold (c) PDMS mixed with the curing compound poured over the mold (d) Curing of the PDMS at 120°C (e) Stripping of the PDMS from the mold (f) Punching of holes in the casted PDMS device, thereby creating fluid connections

A. Fabrication of PMI

In order to make the PMI, a microfluidic mold was fabricated using a silicon wafer (Fig. 9(a)). The wafer was patterned and then time-etched using deep reactive ion etching to achieve the desired channel geometry (Fig. 9(b)). PDMS was mixed with a curing compound (10:1 w/w), poured over the mold, and desiccated to release any trapped air bubbles (Fig. 9(c)). Subsequently, it was cured at 120°C (Fig. 9(d)). After hardening, the PDMS was stripped from the mold (Fig. 9(e)) and punched with two 1 mm holes for inlet and outlet (Fig. 9(f)). The bonding surfaces were cleaned using plasma cleaner PDC-32G (Harrick Plasma Inc.) for 3 minutes followed by the alignment of the channel with the linear array of PMUT cells and bonded to realize the PMI. Post bonding, the PMI was cured at 110°C for 1 hour to ensure the formation of strong bonds.

B. Characterization of PMI

The fabricated channel was characterized using optical profilometer (Talysurf CCI, Taylor Hobson Precision Inc.). The channel depth and width were measured to be 250 μm and ~ 1 mm, respectively, which agreed with the design dimensions. In order to understand the effect of the microfluidic channel integration on the PMUT's properties, vibration responses were obtained before and after the channel bonding. It was found that the channel dampens the vibration from the PMUTs, thereby significantly reducing their quality factor (see Fig.10).

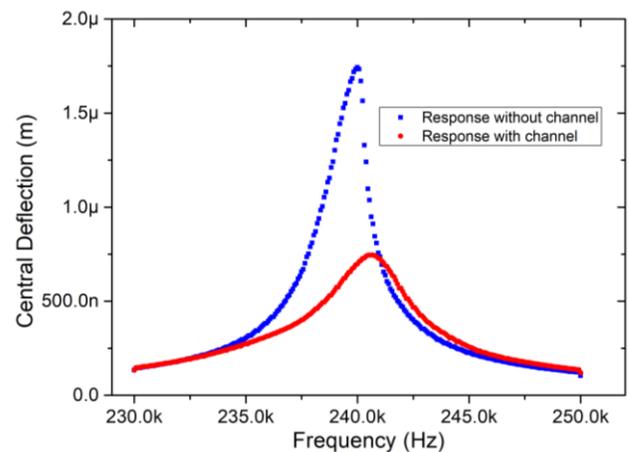

Fig. 10. Effect of channel integration on the resonant response of the PMUT

IV. EXPERIMENTAL SETUP

The PMI was first die bonded to a custom-made printed circuit board using the H70E epoxy from Epotek Inc. and then wire bonded using HB16 wire bonder from TPT Inc. (see Fig. 11). Fluid I/O connections were made by inserting tubings of 1 mm outer diameter into the punched holes. Teflon dissolved in FC-40 (3:7 v/v) was pushed through the microchannel and the PMI was cured at 60°C for 2 hours to establish an insulated teflon coating on the PMUT-fluid interface, thereby electrically insulating the connections from any fluid interference. A lock-in amplifier was connected to the PMI and the input signal

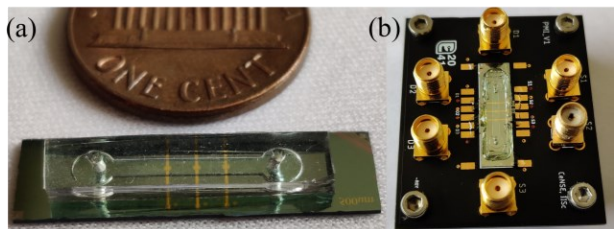

Fig. 11. (a) PMUT Microfluidic Integration (PMI) fabricated using a linear array of three 500 μm diameter PMUTs. (b) PMI wire bonded to a custom-made PCB

frequency was swept in the desired range in order to observe the shift of resonant frequency with the change in density at a driving voltage amplitude of 1V. The tubing at the inlet was connected to a syringe pump for fluid injection. The PMI was isolated from background vibrations using a pneumatic vibration isolation table, and shielded cables were used to minimize electromagnetic interference. The experimental setup is shown in Fig 12.

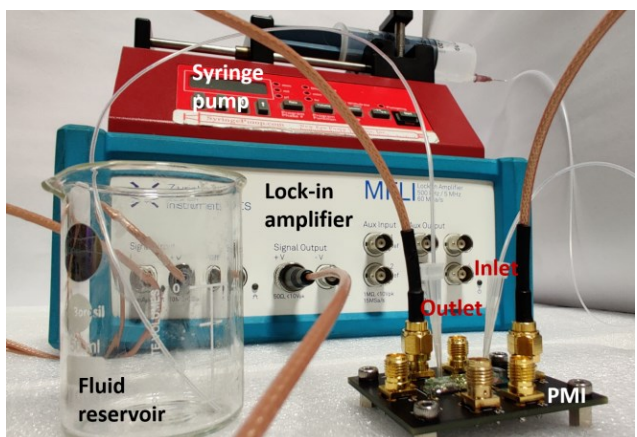

Fig. 12. Experimental setup designed for microfluidic density sensing using PMI

## V. Results

### A. Correlation between the optical and electrical readouts

In order to verify the relation between the deflection caused by the driving signal and the electrical voltage received from the sensing electrodes, a PMUT having 750 μm diameter was driven at 0.1 V to 1 V by sweeping the frequency over the desired range in order to record the maximum deflection using the laser Doppler vibrometer (LDV). Simultaneously, the maximum voltage received from the sensing electrode at the resonance was recorded using the lock-in amplifier. The results so obtained are plotted in Fig 13. It is observed that the maximum deflection obtained from the LDV shows a linear relationship with the voltage received from the lock-in amplifier. The slope obtained from the graph indicates a sensitivity of 0.56 mV/μm.

### B. Density sensing using PMI

Several samples of a test fluid were prepared using ethanol and carbon tetrachloride in ten different solutions in the range of

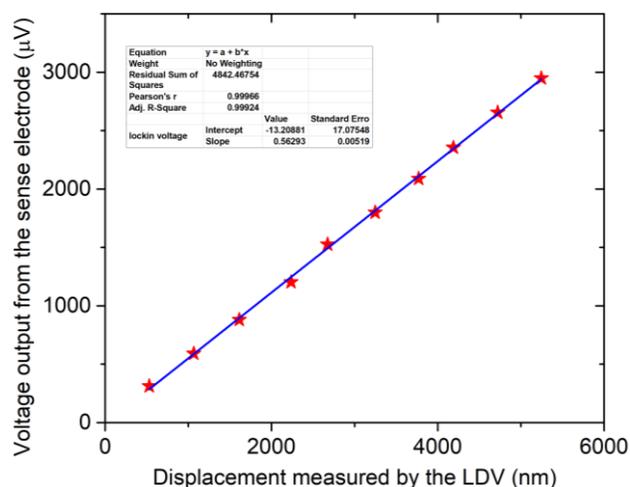

Fig. 13. Maximum deflection obtained from the laser Doppler vibrometer (LDV) compared with the maximum voltage received from the sensing electrode using the lock-in amplifier

774 kg/m$^3$ to 1496 kg/m$^3$ and each of them were individually pushed in the PMI that contained a linear array of three 750 μm diameter PMUTs. The PMUTs were excited and their resonant frequency under the test fluid was found by sweeping the input signal and looking for the peak response obtained from the sensing electrode. The mean resonance frequency is plotted against the test fluid density in Fig. 14. It is observed that the variation of the resonance frequency follows a linear relationship with the fluid density. From the experimental results in Fig. 13, we find the sensitivity of the sensor to be 25.9 Hz/(kg/m$^3$) which follows the same trend as obtained from our previous reported work [19]. The measurements were repeated thrice, and the obtained data was found to be repeatable. The density vs. frequency relationship obtained with a linear fit shows a R-squared value of 0.96, thereby suggesting that PMI can successfully work as a sensor to sense fluid density. Also, the effect of channel depth on the resonant frequency of the PMUTs was simulated. It was found that the resonant frequency shifts by 37%, on varying the channel depth from 10 μm to 1000 μm. After 1000 μm, no significant variation was observed. The simulated results are not included here for the sake of brevity. Thus, in order to select an optimum channel depth, 250 μm was chosen for, in which the resonant frequency differs from the saturated frequency by 6%. Also, the repeatability of the data suggested that a channel of perhaps any reasonable dimension could be used for the resonant density sensing.

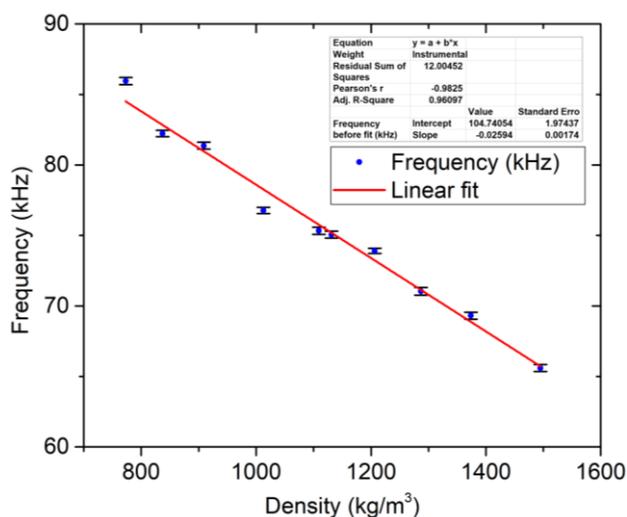

Fig. 14. Frequency vs. density as obtained from the PMUT microfluidic integration (PMI)

### C. PMI as microfluidic blood hemoglobin sensor

In order to evaluate the potential of the PMI as a microfluidic hemoglobin sensor, test fluids were created in the range of 1020 – 1090 kg/m$^3$. This is the range in which the density of human blood varies solely due to the variation in the hemoglobin content [22]. The shifted resonant frequency in each case was recorded and plotted as shown in Fig. 15. The plot shows linear variation of the resonant frequency with the density of the test fluids. The sensitivity of the PMUT as a density sensor is obtained from the slope of the graph which is found to be 26.3 Hz/kg/m$^3$. From the literature, it is found that 1% change in hemoglobin content results in 0.8 kg/m$^3$ change in blood density. Hemoglobin content in human blood generally varies up to 24.6% for males and 24.8% for females [25]. Thus, the sensitivity obtained here corresponds to a density sensitivity of 21.4 Hz/1% change in blood hemoglobin concentration and is considered to be sensitive enough to sense hemoglobin variation in the human body.

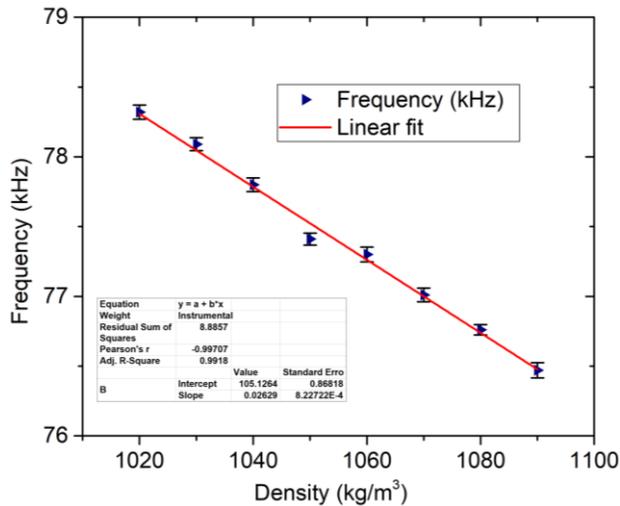

Fig. 15. Frequency vs. Density obtained from the PMUT microfluidic integration (PMI) in the blood hemoglobin range

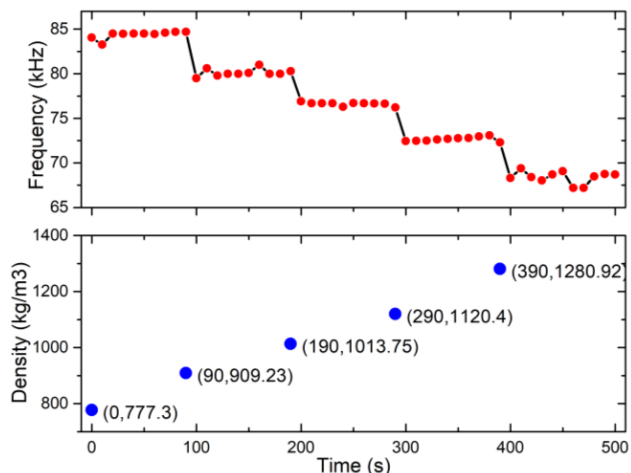

Fig. 16. Real time density sensing using the PMUT microfluidic integration (PMI)

### D. Real-time density sensing using PMI

The performance of the PMI was finally tested for a real-time density sweep in the range of 777.3 to 1281 kg/m$^3$. Carbon tetrachloride was added to ethanol in fixed steps of 100 s and frequency shifts observed after every 10 s using the lockin amplifier and keeping the fluid discharge constant at 1 mL/min. The data obtained is plotted in Fig. 16. The plot shows the variation of the resonant frequency of a typical PMUT in the PMI with time and density of the fluid mixture. It is observed that the frequency switch resulting due to the change in fluid density is almost instantaneous thereby depicting the responsiveness of the PMI towards real time density sensing.

## VI. CONCLUSION

This paper investigates the applicability of a dual electrode, self-sensing PMUT as a fluid density sensor. A unique transceiver platform is created by integrating microfluidics with PMUTs and its capability as a density sensor is demonstrated. It is shown that PMI is capable of detecting fluid densities in a broad range 774 kg/m$^3$ to 1496 kg/m$^3$ with frequency shift in the range of 20 kHz for a typical 750 μm diameter PMUT having a constant stack thickness of 10 μm and negligible residual stresses. The PMI can also sense dynamic change in density with almost instantaneous response, making it an excellent candidate for online fluid density monitoring.

## VII. ACKNOWLEDGEMENT

The authors are grateful to Mr. Veera Pandi N. for his extensive help with PMUT packaging, all the staff at the National Nanofabrication Centre (NNFC), Micro Nano Characterization Facility (MNCF) at the Centre for Nano Science and Engineering (CeNSE), Indian Institute of Science, the funding agencies – the Department of Science and Technology (DST-Nano Mission), the Ministry of Human Resource Development (MHRD), the Ministry of Electronics and Information Technology (MeitY) and all the authors of various literature who have made significant contributions to the scientific community.

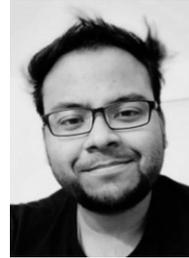

**Kaustav Roy** received his B.E. degree in Civil Engineering from Jadavpur University, Kolkata, in 2016. He is currently pursuing Ph.D. in MEMS Ultrasound at the Centre for Nanoscience and Engineering, Indian Institute of Science. His research interest is experimental ultrasound.

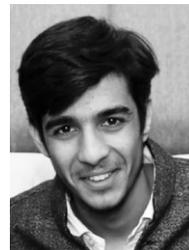

**Kritank Kalyan** received his B.Tech. degree in Mechanical Engineering from Guru Nanak Dev Engineering College, Punjab, in 2018. He is currently working as a project associate at the MEMS lab, Centre for Nanoscience and Engineering, Indian Institute of Science, Bangalore. His research interests include PMUT, microfluidics, droplet impact dynamics and physics of fluids.

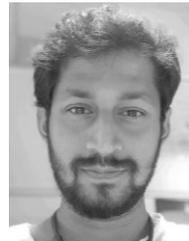

**Anuj Ashok** received his B.E. degree in Mechanical Engineering from Srinivas Institute of Technology, Mangalore, in 2016 and M.Tech. in Nanotechnology from NITK, Surathkal. He is currently working as a project associate at the MEMS lab, Centre for Nanoscience and Engineering, Indian Institute of Science, Bangalore. His research interests include MEMS, VLSI fabrication, ultrasound and photoacoustics.

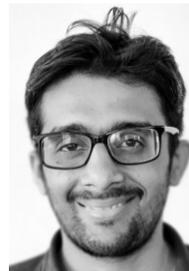

**Vijayendra Shastri** received his B.E. degree in Mechanical Engineering from Sri Jayachamarajendra College of Engineering, Mysore, in 2013. He worked in UTC Aerospace Systems as a stress analysis engineer from 2013-2016. He is currently pursuing Ph.D. in liquid metal flow at microscale at Centre for Nanoscience and Engineering, Indian Institute of Science, Bangalore. His research interests include microscale metal flow and electro-migration

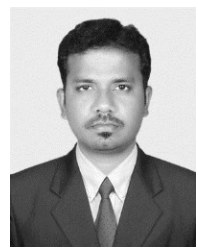

**A. Antony Jeyaseelan** pursuing his Ph.D. in Nanotechnology from CSIR-National Aerospace Laboratories, Bangalore. He worked as a Senior Research Fellow in Materials Science Division, National Aerospace Laboratories, Bangalore. His research interests are sol-gel science, coating technologies, and surface modifications.

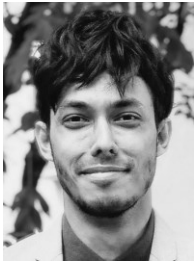

**Avinandan Mandal** received his B.S. from Indian Institute of Science, Bangalore. He worked as a project assistant in the MEMS lab, CeNSE, IISc, Bangalore. His research interests are biosensors, ultrasound and cardiovascular systems.

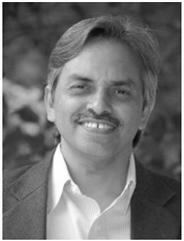

**Rudra Pratap** received his B.Tech. degree from Indian Institute of Technology, Kharagpur, India, in 1985, Master's degree in mechanics from the University of Arizona, Tuscon, in 1987, and Ph.D. in Theoretical and Applied Mechanics from Cornell University, Cornell, NY, in 1993. He is currently a professor in the Centre for Nanoscience and Engineering, Indian Institute of Science, Bangalore, India. His research interests include MEMS design, computational mechanics, nonlinear dynamics, structural vibration, and vibroacoustic.